# Ultrafast disruptive probing: simultaneously keeping track of tens of reaction pathways

Bethany Jochim[1], Lindsey DeJesus[1], and Marcos Dantus[1,2,]*

Ultrafast science depends on different implementations of the well-known pump-probe method. Here we provide a formal description of ultrafast disruptive probing, a method in which the probe pulse disrupts a transient species that may be a metastable ion or a transient state of matter. Disruptive probing has the advantage of allowing simultaneous tracking of the yield of tens of different processes. Our presentation includes a numerical model and experimental data on multiple products resulting from the strong-field ionization of two different molecules, partially-deuterated methanol and norbornene. In combination with ion imaging and/or coincidence momentum imaging, or as complementary to atom-specific probing or ultrafast diffraction methods, disruptive probing is a particularly powerful tool for the study of strong-field laser-matter interactions.

## Introduction

The study of ultrafast phenomena depends on some form of the pump-probe method. The earliest work employed a mirror to reflect the incident pulse in order to create the probe pulse, and the time-resolved dynamics were then recorded onto photographic paper.[1] Sub-picosecond pump-probe methods date back to the mid-1970's, when they were used for observing coherent excitation of molecular vibrations,[2] ultrafast relaxation of electronic states,[3] and the release of CO from carboxyhemoglobin.[4] Multiple approaches for probing physicochemical processes have been developed, some involving photons with wavelengths ranging from the mid-infrared to the X-ray regime, and others involving electron pulses. Here, we describe using a weak probe laser pulse to disrupt product formation initiated by a pump pulse. This method can be used to complement other established probing methods. We discuss situations where disruptive probing can be highly useful, such as in simultaneously following tens of major and minor reaction pathways.

When spectroscopic transitions are well defined, one can select an appropriate probe wavelength to selectively track the dynamics of interest.[5-9] Selective probing has also been accomplished by resonance-enhanced multiphoton ionization (REMPI) using a tuneable laser[10,11,12] or using a fixed wavelength probe pulse and identifying the fragments by photoelectron spectroscopy (PES).[13,14] These two methods work very well when one or two neutral products are expected. Recently, atom-selective probing became possible through monitoring absorption of X-ray photons by inner-shell electrons of atoms with femtosecond and attosecond time resolution.[15-18] In addition to spectroscopic methods, diffractive probing methods such as ultrafast electron diffraction (UED) and ultrafast X-ray diffraction (XRD) have been developed.[19-22]

There are cases where the pump pulse initiates highly complex physicochemical processes with tens or even hundreds of products. Examples of such cases abound in strong-field science.[23,24,25] When intense ultrafast pulses interact with isolated polyatomic molecules, the formation of tens to hundreds of fragment ions and neutral species that are likely vibrationally and electronically hot ensues. Under these circumstances, high-finesse methods like those discussed above are not able to provide a complete account of the strong-field interaction.

A strong-field laser-matter interaction can deposit 10-100 eV of internal energy into a molecule on a timescale that is faster than the motion of the nuclei. Upon strong-field ionization (SFI) of polyatomic molecules, multiple complex chemical reaction mechanisms take place on timescales ranging from femtoseconds to nanoseconds. Despite the large energies involved and the multiple bonds being broken, multiple new bonds may form. Such is the case in the formation of $H_3^+$ from alcohols following double ionization,[26-33] where three bonds break, and three new ones are formed on a 100-350-fs timescale.

Here we discuss an approach that can synchronously monitor tens and in principle hundreds of independent physicochemical pathways, including those with very low yield, with ultrafast time resolution. Our group has been using a similar approach,[29,31,34,35] and more recently, similar measurements have been carried out by others.[36,37] However, this method and how to analyze the data have not been formally discussed in the literature.

[1] Department of Chemistry, Michigan State University, East Lansing, MI 48824, USA
[2] Department of Physics and Astronomy, Michigan State University, East Lansing, MI 48824, USA
* Corresponding author. Email: dantus@chemistry.msu.edu

We refer to this method as "disruptive probing," given that the probe pulse is weak. Therefore, it merely disrupts product formation while it is still occurring. We illustrate this concept using numerical simulations that mimic multiple reaction pathways. Following a brief description of the experimental methods, we show disruptive probing data corresponding to SFI-triggered chemical reactions in partially-deuterated methanol. We also present disruptive probing following SFI of norbornene, a polyatomic molecule with 17 atoms that produces about 60 distinguishable ion signals. We conclude by summarizing the strength of disruptive probing and how it can be complemented by precision probing methods aimed at the selective detection of individual products or pairs of the multiple products.

## The Concept

The SFI process in a polyatomic molecule produces many distinguishable ionized fragments. Using numbers extracted from the NIST mass spectrometry database for alkanes, alcohols, and cycloalkanes, we find that for molecules with more than ten atoms, the number of distinguishable ions is between $1.3n$ and $3.4n$, where $n$ is the number of atoms in the molecule (see Fig. ESI-1 of the Electronic Supplementary Information). This number increases when the molecule has atoms with several stable isotopes like Cl or S. Moreover, SFI often yields doubly- and sometimes triply-ionized species. For SFI of norbornene (17 atoms), we observe about 60 distinguishable fragments, whereas the NIST database reports only 39 fragment ions following 70-eV electron ionization.

For illustration, we consider a generic triatomic molecule ABC, which upon SFI yields the seven singly-charged ions shown in Fig. 1(a). Our goal is to determine the formation time of each product ion. The concept of disruptive probing following SFI is that the bond rearrangement process (cleavage and formation) that leads to the emergence of product ions is preceded by prompt ionization. The initial ionic species is far from equilibrium and has sufficient energy to undergo fragmentation that results in a mass spectrum like that shown in Fig. 1(a). We need a method capable of synchronously interrogating the large number of reaction pathways a polyatomic molecule may exhibit. Here we use a weak probe pulse with polarization at magic angle with respect to that of the pump pulse, capable of disrupting product formation. The probe alone should not be able to ionize or drive multiphoton processes in the molecule. The ion yields when the probe comes before the SFI pump should be unaffected. When the probe comes after the reactions are completed, the ion yields stabilize at their final value. We denote the former and latter cases as "negative" and "long positive" time delays, respectively. Under these conditions, ion yields are dependent on pump-probe delay time from the time of SFI until bond rearrangement processes are complete.

Following SFI one can envision the following scenarios caused by the probe pulse, illustrated in Fig. 1(b): (i) an exponential decay as the ability to disrupt yield decreases with time, (ii) an exponential rise in yield, (iii) the rise and decay of a transient species yield, or (iv) the transient depletion of a final product. We can model these cases by assuming that the Gaussian pump pulse $g(t) = \exp(-t^2/s^2)$ populates ion state $P_1$. Here, $s$ is related to the full width at half maximum (FWHM) pulse duration, by $t_{FWHM} = 2\sqrt{\ln 2}\, s$. The population of $P_1$ decays with time constant $t_1$:

$$\frac{dP_1}{dt} = g(t) - \frac{P_1}{t_1}. \qquad (1)$$

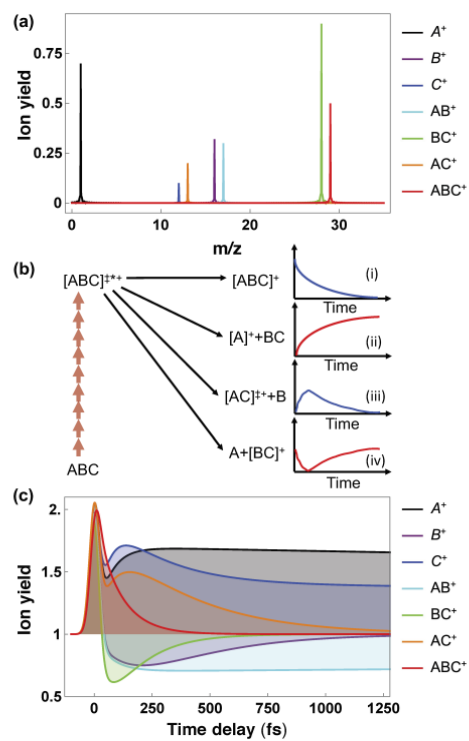

**Fig. 1** The concept of disruptive probing illustrated by numerical simulations. (a) Simulated mass spectrum following SFI of a prototypical molecule ABC showing singly-charged fragments. (b) Different scenarios for time-resolved yields: (i) decay of the precursor ion, (ii) rise of a product ion, (iii) rise and decay of a transient species, and (iv) depletion and recovery of a final product. (c) Disruptive probing of the different fragment ions by depletion (values less than 1) or enhancement (values greater than 1). The strong peak at zero time delay corresponds to the ions resulting from the cross correlation between pump and probe pulses.

Excluding constants, the solution of this equation is

$$P(t,t) = \left[1 + \text{erf}\left(\frac{t}{s} - \frac{s}{2t}\right)\right] \exp\left(-\frac{t}{t}\right). \qquad (2)$$

Function $P(t,\tau)$ corresponds to case (i), while $(1-P(t,\tau))$ corresponds to case (ii). By summing two functions, $aP(t,t_1) + bP(t,t_2)$, one can simulate cases (iii) and (iv), provided their amplitudes have a different sign. When $a = -b$, the result is the same as that used by Koch *et al.* to track the lifetimes of Rydberg states in acetone, which appears as a short-lived enhanced ion yield.[38] A change in the long positive time yield can be modeled by adding another $P(t,\tau)$ term with a very long lifetime. Kinetics, *i.e.*, statistical models are limited in their ability to describe wave packet phenomena in femtosecond experiments, such as oscillating ion yields. This can be modeled by multiplication by cosine function(s) with the appropriate amplitude(s) and phase(s). For data fitting purposes, we added a Gaussian function $g(t)$ at time zero to simulate the cross-correlation signal observed in these experiments. For the numerical examples shown in Figure 1(c), we assume that the probe enhances product formation for fragment ions $A^+$, $C^+$, $AC^+$, and $ABC^+$ and depletes it for fragment ions $B^+$, $AB^+$, and $BC^+$. The parameters used for the simulations and the formula used for fitting the data are given in the Electronic Supplementary Information in Table ESI-1.

Having presented the concept, we address its underpinnings. First, we note that light-matter interactions are greatly enhanced by an induced dipole. Upon SFI, before the electric charge equilibrates and while bonds are lengthening, the ionized transient species $[ABC]^{\ddagger *+}$ is a labile polar radical cation. The disruption by the probe pulse may take place via a one- or two-photon transition to a different potential energy surface, or as a dynamic Stark shift of the potential energy surface responsible for the given reaction, for instance, a barrier suppression. In addition, it has been found that the rate of nonlinear ionization can be strongly enhanced as a molecule is stretched beyond its equilibrium internuclear separation, with the rate being orders of magnitude greater than at equilibrium.[39,40] This type of enhancement may play a role in disruptive probing and explain the transient change in yield of a particular product ion. It is also possible for the probe to ionize a neutral Rydberg state, resulting in enhancement of one or more ion yields. The probe may decrease product yield; for example, in a complex reaction mechanism like a roaming chemical reaction, the probe is likely to mitigate the return of the roaming species and thus prevent product formation. In such a case, product yield is unchanged for negative and long positive time delays. Product yield increases, for example, can occur when an electronegative (electron withdrawing) product that is less likely to end with a positive charge becomes charged because of the probing process.

The magnitude of the change in product yield by the disruptive probe depends on multiple factors like the wave packet speed and spread. In addition, the possibility for a one-photon resonance to nearby states will depend on the wavelength of the probe. The simplest case is when the SFI pump and disruptive probe pulses are the same wavelength. Under these conditions, it is important that the probe pulse have a different polarization than the pump to prevent optical interference. Magic angle (54.7°) or perpendicular polarizations are best. The probe may also be a different wavelength, such as the second harmonic of the SFI pump. In fact, one may be able to confirm the measured reaction times obtained by disruptive probing using different probe wavelengths.

## Experimental methodology

A titanium sapphire chirped-pulse amplification laser system generates linearly polarized ~800-nm, ~40-fs transform-limited laser pulses at a repetition rate of 1 kHz. These pulses are split into pump and probe pulses using a Mach-Zehnder interferometer, and their relative temporal delay is controlled by a motorized linear translation stage. The pump and probe pulses are focused onto a gas target inside a Wiley-McLaren time-of-flight (TOF) mass spectrometer,[41] which facilitates measurement of the resulting positively charged fragment ions. The apparatus and data acquisition are described in further detail in previous publications.[29,35,42]

The focused pump pulses have a peak intensity ~$10^{14}$-$10^{15}$ W/cm$^2$, and the much weaker probe pulses have a typical peak intensity of ~$10^{13}$-$10^{14}$ W/cm$^2$, depending on the system of study. It is key that the pump pulse's intensity is high enough to cause sufficient levels of ionization and fragmentation, while at the same time avoiding saturation to retain sensitivity to disruption by the probe pulse. More specifically, if the pump laser intensity is below that for saturation, $I_{sat}$,[43] then the zero-time delay (temporal overlap of pump and probe) fragment yield spike is very high. If the pump pulse is too intense, it can drive Coulomb explosion of the molecules prior to the reaction of interest, and the temporal overlap yield spike is minimized. Furthermore, a pump peak intensity that is too high can also increase focal-volume averaging effects, which we reduce through use of a 1-mm-wide post-interaction slit in the TOF spectrometer. The slit is orthogonal to the laser propagation direction and selects against ions originating outside the central focal volume.

It is also noteworthy that the pump laser's parameters correspond to the tunneling ionization regime (Keldysh parameter $\gamma$ <1).[44] Some of the high-energy electrons resulting from tunneling ionization can return to re-scatter with the parent molecular ion, and the maximum energy of the returning electrons is approximately $3.2U_p$,[45] where $U_p$ is the ponderomotive energy. These rescattering processes can result in a broad range of internal energies in the molecular ion and can initiate several chemical reactions to be tracked in time.

## Results and discussion

We illustrate the disruptive probing method by presenting data on partially-deuterated methanol, a small polyatomic molecule. The detailed dynamics following SFI depend on the molecule and can be expected to be quite different for different isomers. Disruptive probing results in a rich and comprehensive dataset that allows tracking of numerous reaction products and is thus useful in understanding the fragmentation patterns of polyatomic molecules. We also show data for norbornene, which demonstrates how disruptive probing provides information about all the resulting distinguishable product ions for molecules yielding tens of fragment ions.

**Disruptive probing following strong-field double ionization of methanol**

When methanol molecules are bombarded by high energy (>40 eV) photons,[46] electrons,[47,48] or undergo strong-field ionization,[49] one of the products is $H_3^+$. The formation of $H_3^+$ requires breaking three bonds and forming three new ones. The formation time of the trihydrogen cation from various organic molecules, including methanol, was measured using the disruptive probing method. The results helped reveal that a primary pathway leading to trihydrogen is an unusual roaming mechanism involving a transient hydrogen molecule ejected from a precursor dication.[29,31] Here, we focus on the $H_3^+$, $H_2D^+$ and $H^+$ fragment ions from the $CH_3OD$ molecule. We performed fits to their time-resolved yields using the model based on Eq. 2, as shown in Fig. 2. For the data shown here, the pump's peak intensity was $3\times10^{14}$ W/cm$^2$ ($\gamma \sim 0.6$), and that of the probe was $1\times10^{14}$ W/cm$^2$. The $H_3^+$ fragment ($m/z$=3) from $CH_3OD$ indicates a "local" roaming pathway wherein the roaming $H_2$ originates from the methyl group and later abstracts the remaining proton from the same side of the molecule. One observes in Fig. 2(a) that the $H_3^+$ yield is approximately constant at negative and long positive time delays. In the former case, the weak probe, arriving long before the pump, does not produce any precursor dications and thus has insignificant effect on $H_3^+$ production. In the latter case, the probe also has negligible effect on the already-formed $H_3^+$ ions. A sharp spike near zero time delay indicates the cross-correlation of the pulses. At short positive time delays, the probe causes depletion of the $H_3^+$ ions, which indicates the probe causing fragmentation of the precursor dication before proton capture occurs. With increasing probe delay, the yield recovers to the negative time delay value.

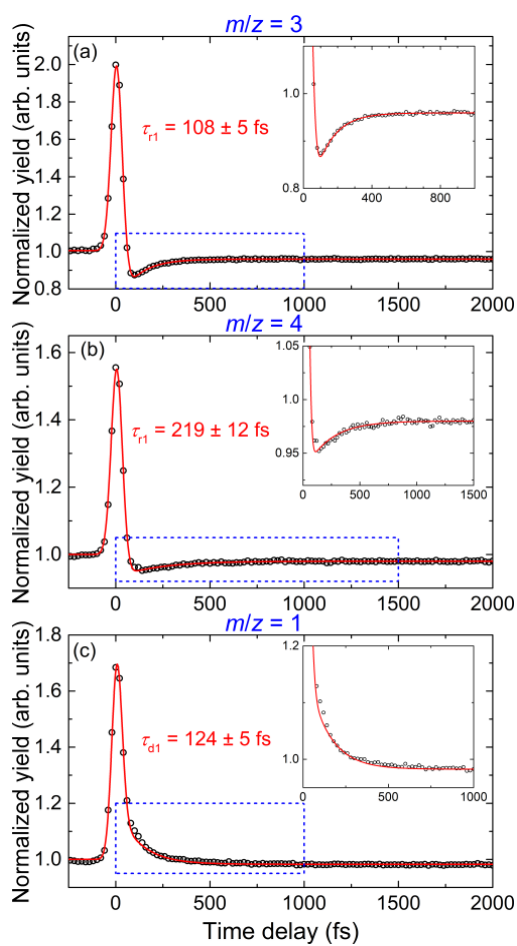

**Fig. 2** Time-resolved yields of fragments from deuterated methanol $CH_3OD$: (a) $H_3^+$, (b) $H_2D^+$, and (c) $H^+$. The data is normalized such that the average yield at negative time delays is unity. Retrieved time constants are indicated in red text, where $\tau_{ri}$ indicates a rise time, and $\tau_{di}$ indicates a decay time. The red lines indicate the fits to the data points. The insets (regions marked with dashed blue boxes) are zoomed in to focus on the dynamics near zero time delay.

This behaviour is consistent with the disruptive probing model case (iv) in Fig. 1(b), allowing retrieval of the reaction time. Using Eq. 2 to fit the data yields the result in Fig. 2(a). This fitting procedure to the whole range of pump-probe delays is an improvement upon our previous exponential fitting solely to the yield recovering after depletion.[29,31] We retrieve an $H_3^+$ formation time of 108 ± 5 fs. Also produced is the $H_2D^+$ fragment ($m/z$=4), which originates from an "extended" roaming mechanism involving two hydrogens from the methyl group and one from hydroxyl side of the molecule. As shown in Fig. 2(b), in agreement with intuition, this mechanism is considerably slower, measured at 219 ± 12 fs. These values are in good agreement with our previous measurements of 119 ± 3 fs and 244 ± 25 fs for $H_3^+$ and $H_2D^+$, respectively.[29,31]

Lastly, we observe a large contribution of proton elimination ($m/z$=1) from methanol, which is quite common for all hydrocarbons. Momentum imaging studies on the strong-field ionization of ethylene found the angular distribution of $H^+$ to be isotropic, implying a process slower than Coulomb explosion.[50,51] In contrast to the depletion and recovery observed for the trihydrogen cations, the time-resolved $H^+$ yield displays an exponential decay with time constant 124 ± 5 fs, as shown in Fig. 2(c). In this case, during the reaction, we observe an enhanced probability for the release of a proton induced by the disruptive probe pulse. Upon double ionization, the formation of neutral $H_2$ is facilitated by the change in electronic configuration of the carbon atom.[29,31] This leaves a metastable $HCOH^{2+}$ species that provides a proton to form $H_3^+$ or simply releases a proton. The first process is diminished by the probe, but the latter is enhanced, with similar timescales.

**Disruptive probing following strong-field ionization of norbornene**

A larger polyatomic molecule best illustrates the strengths of disruptive probing. The SFI of norbornene $C_7H_{10}$ produces about 60 distinguishable fragment ions. The SFI mass spectrum is shown in the Electronic Supplementary Information Fig. ESI-2. The time-resolved data corresponding to 50 of the more prominent distinguishable fragment ions is shown in Fig. 3. Starting from the molecular ion ($m/z$=94) and the corresponding molecule containing one $^{13}C$ ($m/z$=95), we observe fast depletion that does not recover to its original value. Fragment ions $m/z$=77-93 correspond to loss of H atoms/ions. Fragment ions $m/z$=33, 43.5, 46, and 46.5, which do not appear in the NIST electron ionization spectrum, likely correspond to doubly-charged species. The cyclopropenium ion $C_3H_3^+$ ($m/z$=39), which is highly stable, is quite prominent in the mass spectrum.

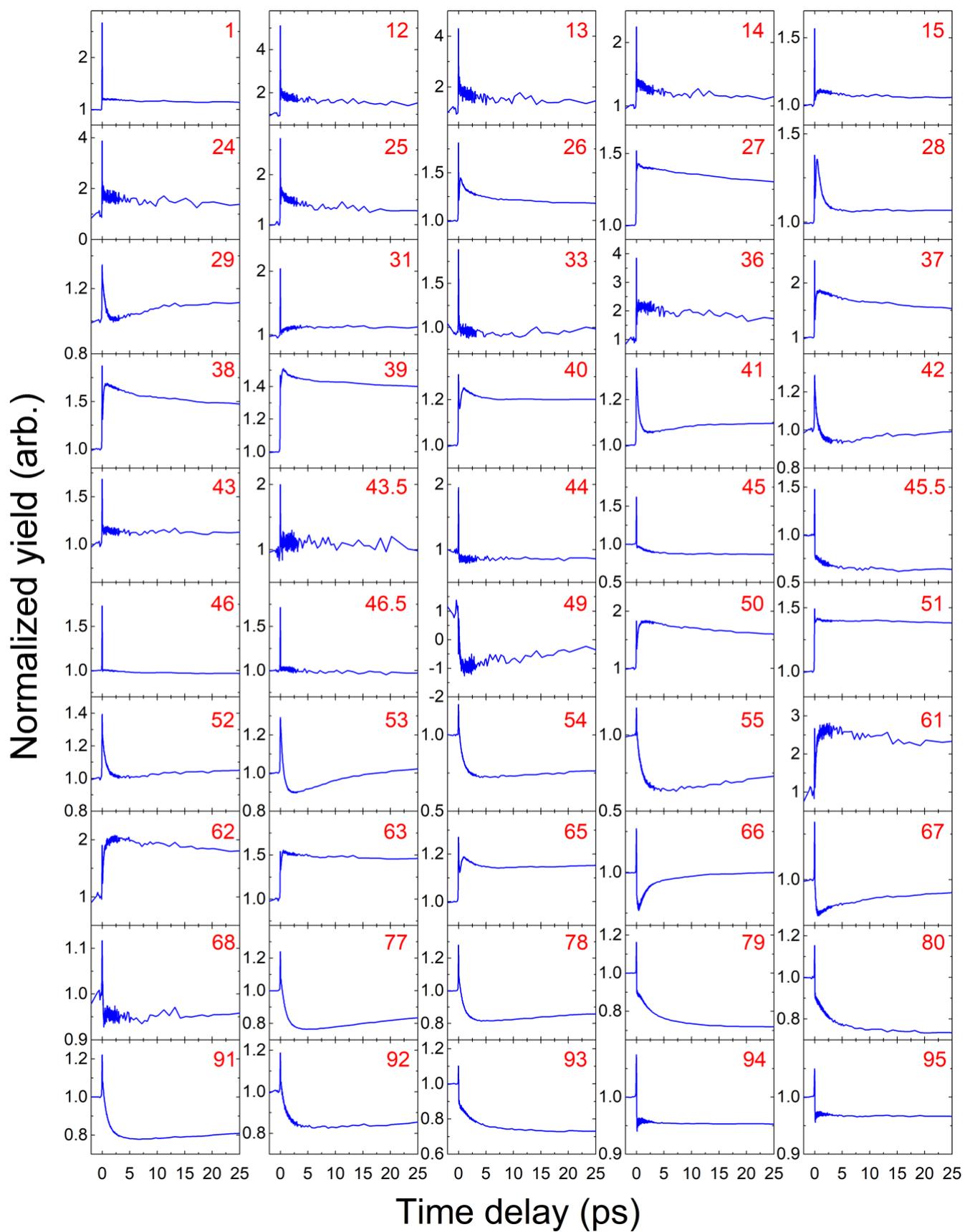

**Fig. 3** Time-resolved yields of 50 distinguishable fragment ions resulting from the SFI of norbornene, m/z=94. The data is normalized such that the average yield at negative time delays is unity. The m/z for each of the fragment ion is indicated by the number in red.

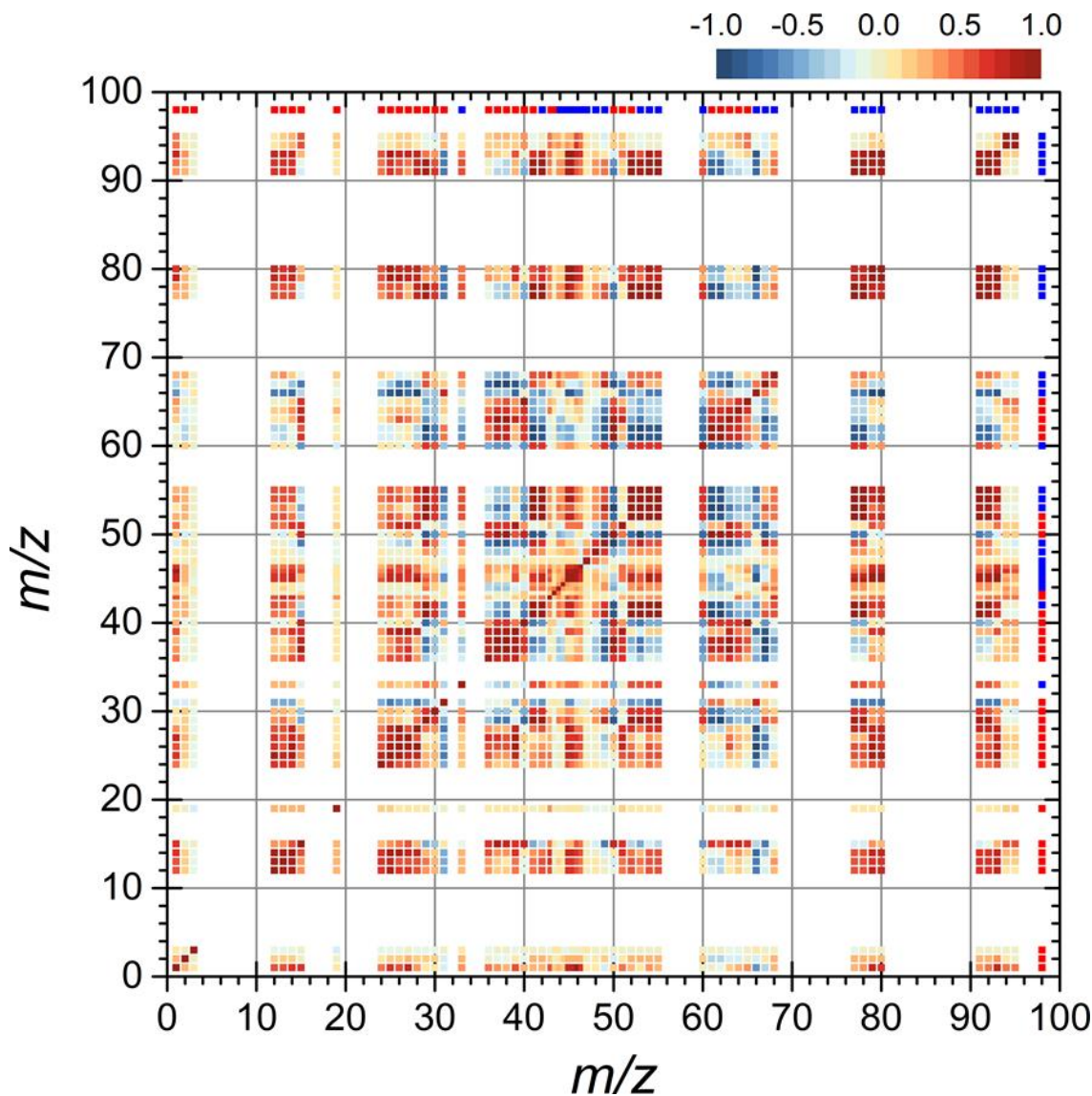

**Fig. 4** Pearson correlation map for 59 distinguishable ions resulting from SFI of norbornene. The numbers on the axes correspond to the *m/z* for each ion. Most transient signals show either a short-lived enhancement or depletion, indicated by the bright red or blue points in row/column 98 at the top and right side of the plot. Enhancement is indicated by bright red, and depletion is indicated by bright blue.

Analysis of tens of different fragment ions can be a daunting task, and thus determining correlation among the fragments can be useful. We have calculated the Pearson correlation coefficient between the time-dependent yields (from pump-probe delay = 40 fs to 50.2 ps) for each possible pair among 59 of the fragment ions. The correlation coefficient $r_{ij}$ for a pair of ions is computed as follows:

$$r_{ij} = \frac{N\left(\sum_{k=1}^{N} Y_{ik} Y_{jk}\right) - \left(\sum_{k=1}^{N} Y_{ik}\right)\left(\sum_{k=1}^{N} Y_{jk}\right)}{\sqrt{\left[N\sum_{k=1}^{N} Y_{ik}^2 - \left(\sum_{k=1}^{N} Y_{ik}\right)^2\right]\left[N\sum_{k=1}^{N} Y_{jk}^2 - \left(\sum_{k=1}^{N} Y_{jk}\right)^2\right]}} \quad (3)$$

Here, $N$ is the number of pump-probe delays. $Y_{ik}$ and $Y_{jk}$ are the yields of ions *i* and *j* at the *k*th time delay.

In the correlation map in Fig. 4, pairs of fragment ions that are positively correlated are indicated by red-brown points, and pairs that are anticorrelated are gray-blue.

The correlation map allows one to quickly determine which fragments have similar time dependences and thus are good candidates for more in-depth exploration. Along the diagonal line, as expected, every fragment displays perfect positive correlation with itself. Close to this line, we find strong positive correlation clusters among compounds differing by one, two, and sometimes three *m/z* units. Note that anticorrelation does not merely indicate that one ion yield increases and the other decreases with time delay; the correlation map is reflective of the timescales of the dynamics. For example, the product from the retro Diels-Alder (rDA) reaction of norbornene, which results in the cyclopentadiene radical cation $C_5H_6^{+\cdot}$ (*m/z*=66),[52,53] is strongly positively correlated with *m/z*=31 and anticorrelated with *m/z*=26, 28, and 39. Anticorrelation of $m/z$=66 with $m/z$=28 may be due to the fact that the rDA reaction preferentially results in neutral ethylene, which can be ionized by the probe if it arrives during the rDA reaction. That is, it is possible that the probe pulse controls which fragment is charged, the cyclopentadiene fragment or the ethylene fragment. Analysis of the correlation between fragments can lead to hypotheses to be tested by more precise approaches, such as coincidence momentum imaging. For example, the probe transiently depletes the yield of the product with $m/z$=66 but enhances the yield of $m/z$=39, with minima/maxima at similar time delays. A future experiment utilizing coincidence momentum imaging techniques could allow delving into the specific channel(s) producing these fragments in further detail. We now focus on the dynamics of a few of the most important product ions. Of particular interest is $m/z$=66, whose ion yield displays oscillations with a damping constant of ~2.5 ps, as shown in Fig. 5(a). The oscillations correspond to a vibrational mode of the norbornene ion that is likely excited upon ionization from neutral ground state norbornene, and its motion promotes the rDA concerted reaction.

In Figure 5(b), we show the time-dependent behavior of $m/z$=26 (the acetylene ion $C_2H_2^+$). The acetylene ion displays behavior that is anticorrelated with that of $m/z$=66. During the retro-Diels-Alder reaction, the positive charge tends to be on the larger fragment $C_5H_6^+$, and the remaining ethylene fragment is neutral. Disruptive probing can cause ionization and $H_2$ elimination from ethylene to form acetylene,[50,54] thus leading to an enhanced yield that is even more probable before vibrational cooling.

Finally, we present in Fig. 5(c) $m/z = 50$ ($C_4H_2^+$). This fragment is the diacetylene cation,[55] which has been observed following electron ionization of norbornene.[52] This molecule is prevalent in combustion and responsible for a diffuse interstellar band.[56]

## Conclusions

In this study, we have presented a formal description of disruptive probing and its ability to simultaneously follow tens of different pathways. We provide a kinetic model that can be used to simulate and fit data resulting from this type of study. We have also outlined the experimental requirements to best implement this method.

As stated in the introduction, there are numerous probing techniques that provide highly specific information based on known spectroscopic transitions. Similarly, diffractive techniques can follow the entire molecular structure as it undergoes a chemical reaction. However, following high-energy excitation via electrons, photons, or SFI, the target molecule may undergo fragmentation along multiple pathways on complex multidimensional potential energy surfaces. Under these conditions, disruptive probing, as presented here, produces a rich set of data that can then be further explored by complementary approaches.

The experimental data presented on partially-deuterated methanol and norbornene illustrate the power of this method. For norbornene, we have shown the time-dependent ion yields for 50 of the main fragment ions resulting from SFI. We also investigate fragment ions having closely correlated or anticorrelated dynamics, an analysis that can be used to find precursors and product relationships among ions. In the cases of low correlation, one can determine that a pair of ions are formed via very different reaction pathways.

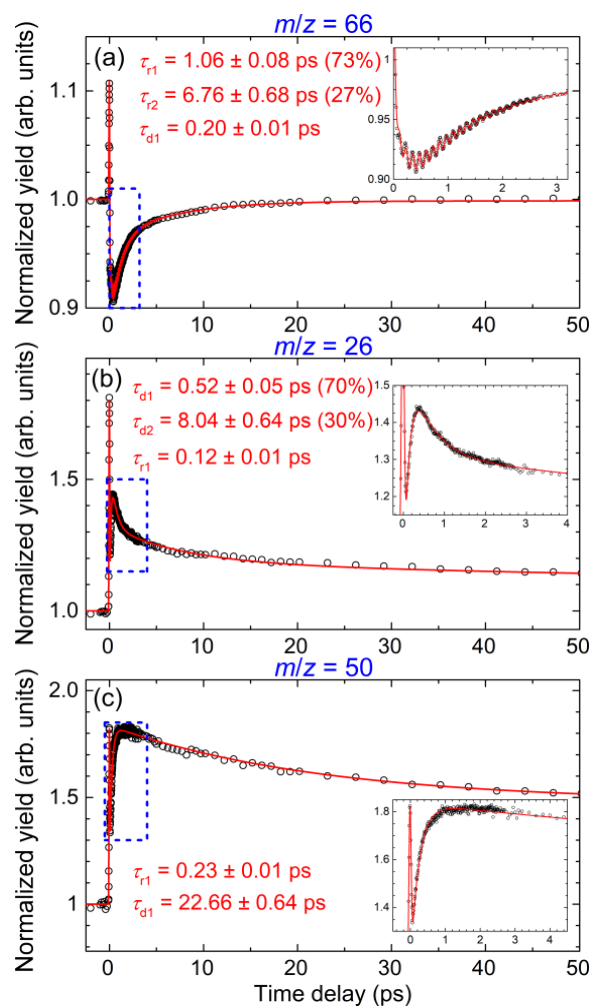

**Fig. 5** Time-resolved yields of fragments from norbornene: (a) $C_5H_6^+$ (b) $C_2H_2^+$ and (c) $C_4H_2^+$. The yields have been normalized in the same manner as those in Fig. 2. Fit functions are indicated by the red lines, with retrieved parameters shown in red font. In the cases of two retrieved rise or decay constants, the numbers in parentheses indicate the percent contribution of that component. The insets are zoomed in on the regions in the dashed blue boxes to highlight behavior at short delays.

Finally, we present a detailed analysis of three different fragment ions from the SFI of norbornene. In addition to the product ion associated with the rDA reaction, we present the formation of the acetylene and diacetylene cations. We illustrate how the model provided in the introduction allows satisfactory fits to the data, resulting in time constants that can be associated with the chemical reaction times.

We consider disruptive probing to be an important tool for ultrafast sciences, one that can be extremely useful when applied with various ion and/or electron measurement techniques. By applying the disruptive probing scheme along with momentum imaging or coincidence momentum imaging, one gains additional specificity in terms of the origin of the fragments. For example, this is how it was determined that $H_3^+$ from methanol comes predominantly from dissociative double ionization and not from single ionization.[29,31] Furthermore, the detailed examination of fragmentation channels at the level of kinetic energy release and angular distributions that these techniques facilitate, together with disruptive probing, could be quite powerful.

## Conflicts of interest

There are no conflicts to declare.

## Acknowledgments

We gratefully acknowledge Dr. Shuai Li for helpful discussions about fitting the data. This material is based upon work supported by the U.S. Department of Energy, Office of Science, Office of Basic Energy Sciences, Atomic, Molecular and Optical Sciences Program under Award Number SISGR (DE-SC0002325).